# Stability by Design: Atomistic Insights into Hydrolysis-Driven MOF Degradation

Ashok Yacham,[†] Tarak K. Patra,[†,$] Jithin John Varghese,[†] Richa Sharma[‡*]

[†]Department of Chemical Engineering, Indian Institute of Technology Madras, Chennai 600036, India

[$]Center for Atomistic Modeling and Materials Design, Indian Institute of Technology Madras, Chennai 600036, India

[‡]Massachusetts Institute of Technology, Cambridge, Massachusetts 02139, USA

*Corresponding author: richas@mit.edu

**Abstract**

Metal-organic frameworks (MOFs) are porous materials formed by interconnected metal atoms via organic linkers, resulting in high surface area and tuneable porosity, making them exceptional candidates for $CO_2$ capture. However, their stability and efficacy in humid conditions are not fully understood, often limiting their commercial applications. Here, we estimate the stability of seven common Zn-based MOFs using reactive molecular dynamics (MD) along with metadynamics sampling to determine hydrolysis energetics at conditions representative of low water concentration limit. The reactions' free energy surfaces (FESs) showed that water stability strongly depends on its linker size and chemistry. Our findings indicate zeolitic imidazolate frameworks (ZIFs), a subclass of MOFs, exhibit higher water stability than iso-reticular metal-organic frameworks (IRMOFs). We further attempt to correlate hydrolysis energy barrier with the physicochemical descriptors of these MOFs. This study provides insights into the critical factors and fundamental implications for developing stable porous materials for carbon capture technologies.

**Introduction**

The chemical absorption of $CO_2$ using amine solvents is projected to be the leading technology for large-scale carbon capture from point sources. However, this approach faces significant challenges, including chemical degradation, corrosion of capital-intensive equipment, excessive water and energy usage, and amine emissions. As a result, sorbent technologies utilizing novel materials are gaining attention. Among these, metal-organic frameworks (MOFs) could represent the future of adsorption-based carbon capture. MOFs are porous and crystalline structures composed of metal ions linked by organic ligand molecules, exhibiting high surface area[1]. Their unique structural properties allow for tuneable porosity and surface area, making them highly effective at selectively capturing $CO_2$ from various sources. Amidst ongoing advancements in large-scale synthesis, these materials have been explored to separate $CO_2$ and $H_2$ storage applications. MOFs such as Zn-MOF-74 (Zn-dioxide benzene dicarboxylate), IRMOF-1(Zn-benzene dicarboxylate), and MOF-177 (Zn-benzene tribenzoate) have exceptional carbon dioxide loadings as high as 19.8 wt%, 8.5 wt%, and 6.5 wt% at ambient temperature (293-313 K) and low pressure (<1.2 bar) [2]. At 77 K, IRMOF-1 has a hydrogen storage capacity of 5.1 wt% at 80 bars, while MOF-177 has a slightly higher capacity of 7.5 wt% at 70 bars, making them promising materials for hydrogen storage[3,4]. One of the major concerns of MOF based technologies is that many MOFs suffer from thermal, mechanical, and moisture-induced chemical degradation, limiting their practical applications. As such the stability of materials in wet or humid environments is a critical determinant of their practical utility across a wide range of applications, including gas adsorption, catalysis, sensing, and separation[5,6]. In many of these applications, materials are exposed to water vapor or liquid water either as part of the operating environment or during regeneration cycles. For instance, in gas separation or storage, water molecules may compete with target gas species for adsorption sites, or even degrade the framework by breaking metal–ligand coordination bonds, leading to loss of crystallinity and porosity. Similarly, in catalysis, especially under aqueous-phase or humid conditions, the structural integrity of the catalyst is essential to maintain its activity and selectivity over time. In sensing applications, exposure to moisture can not only influence the sensing mechanism but also result in irreversible damage if the material is not hydrolytically stable. Therefore, understanding and improving the water stability of materials—particularly porous coordination frameworks like MOFs—is fundamental for translating their promising laboratory-scale properties into robust, real-world technologies. This includes investigating the role of metal-ligand bond strength, hydrophobicity of the framework, pore chemistry, and possible post-synthetic modifications aimed at enhancing hydrolytic resistance.

Akiyama et al. and Wade et al. used partition coefficients as descriptors to represent hydrophobic nature corresponding to water adsorption behaviour of MOFs[7,8]. Especially, the correlations between structural stability with the humidity are not well understood. While it is known that exposure to moisture can impact the integrity and performance of MOFs, the precise mechanisms governing these effects at the atomic level are still unclear. Understanding how humidity influences framework decomposition, pore collapse, or potential structural rearrangements is essential for optimizing MOF design for applications in gas storage, separation, and catalysis. Recent efforts have been directed at enhancing the stability of these frameworks, as their resilience is vital for maintaining structural integrity in flue gas environments. It appears that most existing MOFs show limited endurance in humid conditions, significantly hindering their effectiveness for carbon capture, especially given the high humidity levels commonly present in flue gas. The typical composition of flue gas from a coal-fired power plant usually contains 5-7 % by volume of water vapor[9]. So, it is evident that when a humid flue gas was treated for carbon dioxide separation using MOFs as adsorbents, an adsorption site would experience competition for both $CO_2$ and $H_2O$. So far, CALF-20 has been reported as a stable material for up to 450,000 cycles of steam treatment [10]. The moisture in MOF could lead to structural deformation and either increase or decrease the $CO_2$ capacity [11,12,13,14]. Experiments and molecular dynamics have confirmed that IRMOF-1 decomposes beyond 4 wt% of water content [15,16]. Similarly, MOF-177 degrades with water [17], whereas ZIF-8 has exhibited high thermal stability and chemical resistance to organic solvent and boiling water [18]. Investigation on ZIF-68, ZIF-8, ZIF-90, ZIF-Cl, ZIF-$NO_2$, and SALEM-2 for $CO_2$ capture in the presence of water vapor and sulphur dioxide showed that selectivity for $CO_2$ increases in the presence of water ($H_2O$) and sulphur dioxide ($SO_2$) in polar functionalized ZIFs [19,20]. Building upon these prior research, here, we explore the interactions between MOFs and water at the atomic scale, providing a detailed analysis of their reaction pathways. By investigating the underlying mechanisms and energy landscapes, we aim to elucidate how water molecules engage with MOF structures and influence their stability.

The MOF-water reaction can occur through two mechanisms: i) ligand displacement and ii) hydrolysis. Inserting a water molecule at a metal-ligand bond vicinity would result in undissociated water interacting with the metal and a dangling linker, i.e., ligand displacement, while water dissociates during hydrolysis, resulting in metal-ligand bond breakage, forming metal hydroxide and a protonated ligand[21]. A stability region map was presented by Canivet and co-workers showing MOFs exposed to the variable steam and temperature with their corresponding activation energy for ligand displacement by water[22]. A reaction force field (ReaxFF) based molecular dynamics (MD) tool was utilized by Han et al. and Liu et al.[23,24] to predict the critical water content that causes deformation in IRMOFs at a temperature of 300 K. In addition, the activation barrier for the dissociation reaction of a single water molecule in Zn-MOF-74 was reported as 22.5 kCal/mol using ReaxFF MD[24] and DFT[25], while for two water molecules, it was 15.2 kCal/mol [24]. Using DFT method a similar trend was observed by Zuluaga et al.[26]. This reduction in the barrier was not intuitive, given two metal-ligand bonds must be broken during hydrolysis. Recently, Yang et al. [27] combined ReaxFF based MD simulations with biasing techniques such as metadynamics to investigate the stability of ZIFs and MOFs in the presence of water. MOFs with larger pores are less stable, while ZIFs with $ZnN_4$ exhibit higher aqueous stability. As the water dissociation, particular at the low concentration limit, is a rare event and challenging to capture within a standard MD simulation timeframe, the metadynamics is used to accelerate the process[28,29].

Here, we adopt this strategy of ReaxFF-based MD simulations and metadynamics to establish the free energy surface (FES) of MOF dissociation reaction for seven well-known MOFs – IRMOF-1, IRMOF-10, IRMOF-14, MOF-177, Zn-MOF-74, ZIF-4 and ZIF-90. By selecting the right set of collective or control variables and applying constraints, we overcome the difficulty of visiting energetically separated meta-stable configurations. Our controlled variables-driven metadynamics simulations provides insights to the dynamic evolution of the system over time, showing the reaction pathways and enabled us to estimate the underlying free energy surface (FESs). We also derive insights into the contributions of organic ligands to the stability of these MOFs. We estimate the timescale of hydrolytic degradation of these MOFs for very low humidity conditions. We expect these findings will be useful in the design of stable and efficient materials for carbon dioxide capture in wet flue gas applications.

1. **Model and Methodology**

The ReaxFF-based MD and metadynamics simulations were implemented using the Software for Chemistry and Materials (SCM), Amsterdam Modelling Suite (AMS 2023.102) [30]. Since majority of the Computation-Ready Experimental Metal-Organic Framework Database (CoRE MOFs 2019)[31] reported are zinc-based materials, we focused on representative Zn-based MOFs, including those containing benzene dicarboxylate (BDC), biphenyl dicarboxylate (BPDC), pyrene dicarboxylate (PyDC), benzene tribenzoate (BTB), dihydroxy benzene dicarboxylate (DOBDC). Additionally, we explored a subclass of MOFs called ZIFs with imidazolate (Im) and iso-carboxaldehyde (Ica-Im) ligands. The specific MOF frameworks investigated were IRMOF-1 ($Zn_4O(BDC)$), IRMOF-10 ($Zn_4O(BPDC)$), IRMOF-14 (($Zn_4O(PyDC)$), MOF-177 (($Zn_4O(BTB)_2$), MOF-74 ($Zn_2(DOBDC)$), ZIF-4 ($Zn(Im)$) and ZIF-

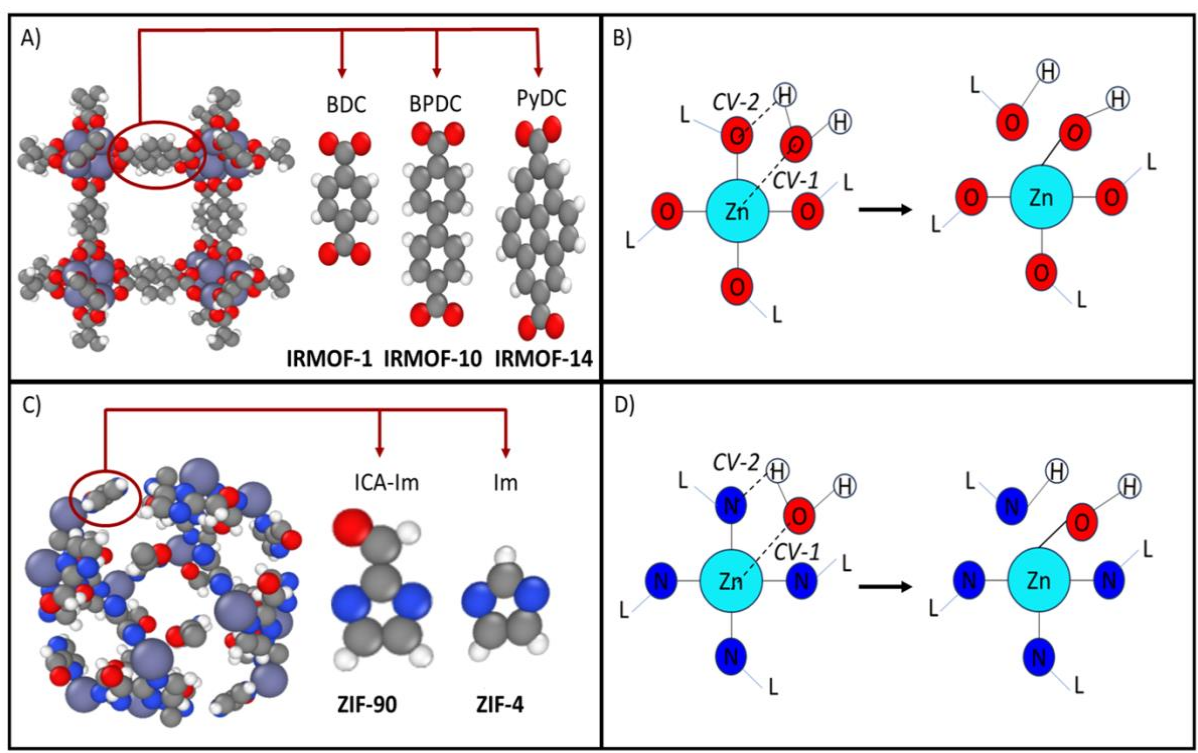

*Figure 1: A unit cell of an IRMOF structure is shown in (A). The linker position between metal atoms is encircled. Three types of linkers -BDC, BPDC and PyDC, which correspond to IRMOF-1, IRMOF-10, and IRMOF-14, respectively, are schematically shown in (A) as well. A unit cell of ZIF is shown in (C). The linker position in ZIF is encircled. Two types of linkers – ICA-Im and Im, which correspond to ZIF-90 and ZIF-4, respectively, are schematically shown in (C) as well. Color-coded spheres represent atoms: grey, red, white, and blue (small and large) C, O, H, N, and Zn, respectively.* **Schematic representation of hydrolysis reaction in B) IRMOF and D) ZIF.** *Key control variables, including the distances between the metal and water oxygen (Zn-Ow) and the linker oxygen/nitrogen and water hydrogen (Oc/Nc-Hw), are designated as CV-1 and CV-2. The spheres are color-coded atoms: cyan, red, blue, and white color spheres represent Zn, O, N, and H, respectively, with L representing the organic ligand that connects the metal centres.*

90 ($Zn(Ica-Im)$). The differences in the building of IRMOFs and ZIFs, marking the metal-organic ligand bond, are illustrated in Figure 1. Crystallographic information files (cifs) for all these materials were obtained from the Cambridge crystallographic data center (CCDC) [18,32,33]. Initial unit cell optimization was performed using the ReaxFF parameters from Yang et al. [27], achieving an energy accuracy of 1E-05 Ha. To determine the reactivity of the MOF with water, we positioned water at a distance of approximately 5 Å from the targeted Zn metal atom in close proximity to the linker. The MD/metadynamics simulations were initiated from this initial configuration to study the hydrolysis reaction. The proposed steps for the hydrolysis reaction are as follows: i) a water molecule approaches the Zinc (Zn) atom, and ii) the water molecule dissociates to form hydrolysed products; OH with the Zn metal site, displacing the linker-Zn bond and creating a dangling linker as shown in Figure 1B, 1D. Simultaneously, the nitrogen (N) or oxygen (O) atom in the disconnected linker abstracts a hydrogen atom (H) from the water molecule, forming an NH or OH species.

We used two collective variables: (i) CV-1: the distance between one of the proximal Zn atoms in the MOF and the oxygen atom of the water molecule (denoted as Zn-Ow) and (ii) CV-2: the distance between the linker oxygen atom and hydrogen atom of the water $O_c$-Hw for IRMOFs as shown in Figure 1B, and $N_c$-Hw for ZIFs as depicted in Figure 1D. These collective variables are crucial for capturing the reactive events leading to initial defect formation in the MOFs. Using the PLUMED library, [34] we constrained the control variables with quadratic upper walls, applying a bias factor of 100. Gaussians of 5 kJ height and 0.1 Å width were filled at a frequency of every 1000 MD steps. The MD simulations were performed in a canonical ensemble (NVT) using an integration time step of 0.25 femtoseconds for a duration of 3 nanoseconds to study the interaction of water with the MOFs. The Nose-Hoover [35] thermostat was used to maintain the system's temperature at 300 K, with a coupling constant of 100 femtoseconds. The PLUMED tool generated metadynamics output in terms of potential as a function of the control variables, which was then used to construct the two-dimensional FES plot. To determine the adsorption energy of water molecule in MOFs, a molecular dynamics simulation in an NPT ensemble was performed for 0.5 ns with a timestep of 0.25 fs at 300 K and 1 bar without any constraints. The adsorption energy was estimated using the standard expression presented in the SI.

## 2. Results and Discussions

We first optimized the geometries of 7 MOF structures using the ReaxFF. The optimized lattice parameters are compared with the experimental data in Table 1. Our simulation structures are in close agreement with experimental findings reported in the literature. We also note that our lattice constant calculations are in agreement with previous computational studies[23,36]. Subsequently, we perform a MD-metadynamics study with these optimized structures.

*Table 1: Calculated lattice constants are compared with experimental reports. All the constants are in Å. The simulation data are based on the ReaxFF.*

| MOF | Simulations | Experiments |
| --- | --- | --- |
| IRMOF-1 | a=b=c=26.7582 | a=b=c=25.8320 [37] |
| IRMOF-10 | a=b=c=35.333 | a=b=c=34.2807 [37] |
| IRMOF-14 | a=b=c=35.2703 | a=b=c=34.381 [37] |
| MOF-177 | a=b=37.904, c=30.8056 | a=b=37.0720, c=30.033 [38,39] |
| Zn-MOF-74 | a=b=26.8933, c=7.1592 | a=26.001, c=6.8271 [40] |
| ZIF-90 | a=b=c=16.98 | a=b=c=17.25 [32,41] |
| ZIF-4 | a=15.5464, b=15.354, c=18.6902 | a=15.3950, b=15.3073, c=18.4262 [18] |

The evolution of the collective variables in the constructed FES of IRMOF-1 hydrolysis is shown in Figure 2. Initially, the water molecule is positioned at a distance of around 3.5 Å from the Zn site. The water molecule approaches the Zn centre multiple times before moving away within the first 600 ps, as shown by the Zn-Ow distance in Figure 2A. The initiation of hydrolysis is observed around 450 ps when the Oc-Hw distance is around 1 Å, as shown in Figure 2B. Hydrolysis is deemed complete when the Zn-Ow distance is approximately $2 \pm 0.1$ Å, and the Oc-Hw distance is around $1 \pm 0.1$ Å, as depicted in Figures 2A and 2B, respectively. The FES reveals a stable complex of undissociated water interacting with the MOF at approximately 2.3 Å from the Zn centre. The corresponding atomic structure is shown in Figure 2C as the reactant and is followed by the dissociated and the hydrolysed complex as product. Although metadynamics simulations do not reveal a saddle point or transition state, a minimum energy path connecting the undissociated interaction complex and the hydrolysed complex is illustrated by the red dashed curve in Figure 2C. The highest energy point along this curve approximates the transition state for hydrolysis, the corresponding atomic structure is shown in Figure 2C as the transition. The activation-free energy barrier for the hydrolysis of IRMOF-1 was estimated to be $9 \pm 1$ kCal/mol (Table 2) with a free energy change of hydrolysis of 0 kCal/mol (Figure 2C). We note that Low et al. [21] reported about 11 kCal/mol of activation energy for ligand displacement by water vapor in IRMOF-1.

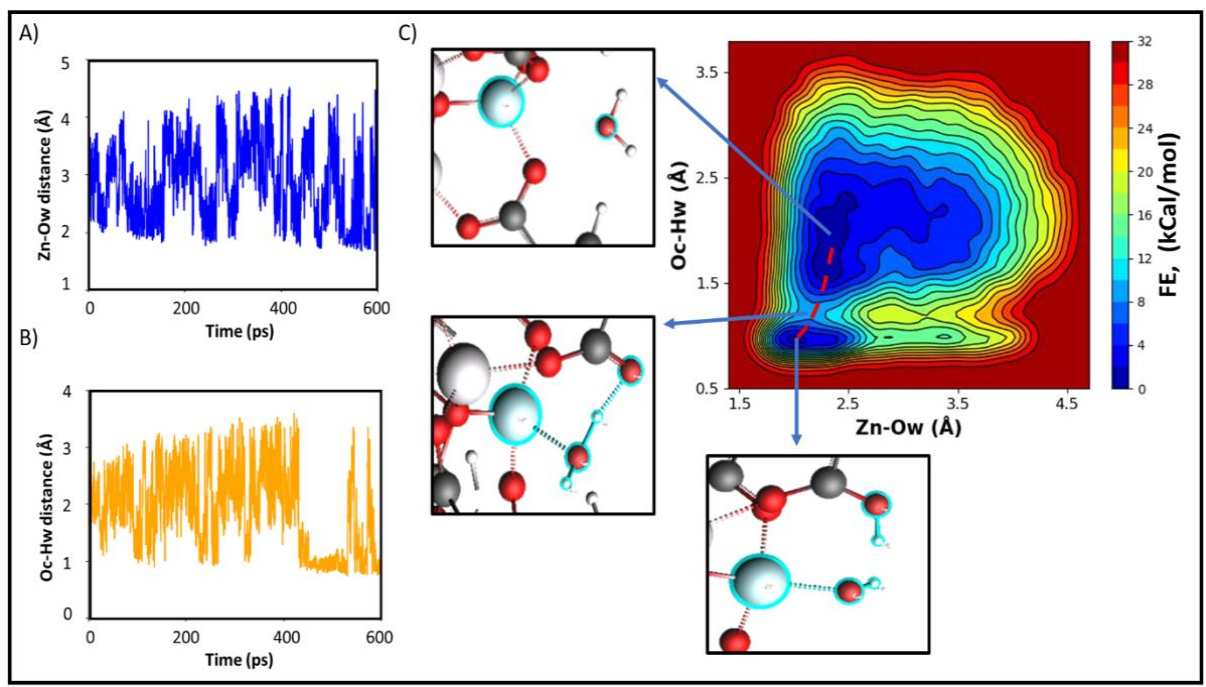

*Figure 2: Progression of the collective variables during a metadynamics simulation of IRMOF-1 (Zn-BDC) with a water molecule at 300 K. A) The distance from the IRMOF metal Zn site to the oxygen of water (Ow) is shown as a function of simulation time, B) The distance from the linker oxygen (Oc) to water hydrogen (Hw) is shown as a function of simulation time. Free energy surface as a function of these two collective variables, constructed from the metadynamics simulation over 600 picoseconds, illustrating hydrolysis of IRMOF-1, is shown in (C). Atomistic representation alongside the FES profile highlighting the reactant to product formation connecting path on the FES. Spheres coloured grey, red, and white (small and large) represent C, O, H, and Zn respectively.*

While IRMOF-10 and IRMOF-14 exhibit broadly similar trends as shown in SI (Figures S1, S2), several distinct differences become apparent upon closer inspection of the trajectory and the FESs. In the case of IRMOF-10, the frequency with which the water molecule approaches and retreats from the Zn centre (Figure S1A) is noticeably higher than in IRMOF-1 (Figure 2A), indicating that IRMOF-10 is slightly less accessible to water than IRMOF-1. The initiation of hydrolysis begins around 500 ps, when the Oc-Hw distance becomes approximately 1 Å, as indicated in Figure S1B. This marginally longer time scale indicates that the hydrolysis process in IRMOF-10 is slightly more challenging than in IRMOF-1. The stable interaction complex of the undissociated water with the MOF has a slightly increased Zn-Ow distance of 2.5 Å, in Figure S1C, further confirms the reduced accessibility of the Zn site to the water molecule in IRMOF-10, as noted earlier. These factors contribute to an increased activation-free energy barrier for the hydrolysis of IRMOF-10, estimated to be 15±1 kCal/mol (Table 1), with a free energy change for hydrolysis of 0 kCal/mol (Figure S1C). IRMOF-14 exhibits an even lower accessibility to water molecule, marked by significantly more frequent events of the water molecule approaching and retreating from the Zn centre, as shown in Figure S2A. Hydrolysis initiation occurs at a notably longer timescale of 1 ns (Figure S2B). The estimated activation-free energy barrier for this process is notably higher at 19±1 kCal/mol (Figure S2C), indicating that hydrolysis is considerably less favourable, with a free energy change of hydrolysis of 4 kCal/mol (Figure S2C). The results suggest systematic trends linked to the nature of the organic ligands in IRMOF-1, IRMOF-10, and IRMOF-14. From IRMOF-1 to IRMOF-10 and IRMOF-14, the number of carbon atoms in the organic ligand increases. This increase highlights the significance of the carbon-to-oxygen ratio, which appears to have a linear correlation with the activation-free energy barrier for the hydrolysis reaction. Our Metadynamics simulations reveal that the activation free energy barrier for the hydrolysis step rises with the organic ligand's carbon-to-oxygen ratio, as shown in Figure 4C. This characteristic ratio drives the hydrophobic behavior of MOFs in the presence of water. The larger the carbon-to-oxygen ratio, the higher the energy penalty for hydrolysing the organic ligand. Building units with higher carbon content in organic ligands creates larger pores, enhancing water repellence in IRMOFs (Table 1). Thus, based on our simulation results, the carbon-to-oxygen ratio could serve as a vital screening factor and design criterion for developing water-stable IRMOFs for wet flue gas capture applications.

Next, we calculate FES of ZIF-90 hydrolysis. In IRMOFs, the metal centres coordinate with the oxygen atoms of the carboxylic acid groups in the organic ligands, as shown in Figure 1A. In contrast, in ZIFs, the metal atoms are coordinated with the nitrogen atoms of the imidazole ring, as shown in Figure 1B. The adsorbed water molecule in MOFs can dissociate to form hydroxyl group and proton, which target the metal-ligand covalent bond, resulting in the formation of Zn-OH and linker-H bonds (O-H in IRMOFs and N-H in ZIFs as shown in Figure 1C and 1D). The metadynamics simulation of hydrolysis of ZIF-90, which contains 2-imidazole carboxaldehyde as the linker, reveals both similarities and notable differences compared to previously discussed IRMOFs. The lifetime of the water molecule near the Zn site is notably short, as it frequently moves away and returns (Figure 3A). This behavior indicates limited accessibility of water molecule to the metal site, similar to what is observed in IRMOF-10 and IRMOF-14. This can also be seen from the shift of the interaction complex of water with the metal site to a distance of around 3.1 Å (Figure 3C). The hydrolysis reaction in ZIF-90 initiates over a significantly longer timescale of around 2200 ps (Figure 3B), in contrast to IRMOF-1, where initiation occurs around 450 ps. This is also reflected in a much higher activation-free energy barrier for the hydrolysis of 33±1 kCal/mol in ZIF-90 (Figure 3C) compared to 9±1 kCal/mol in IRMOF-1. Additionally, the reaction is, highly unfavorable with a reaction-free energy of 28 kCal/mol (Figure 3C), in contrast to the more favourable 0 kCal/mol seen in IRMOF-1. These results clearly indicate that ZIF-90 is significantly more stable against water attack than IRMOFs, which is consistent with experimental observations. Thus, the nature of the metal-ligand bonding and local coordinating significantly influences the stability of the MOF against water attack.

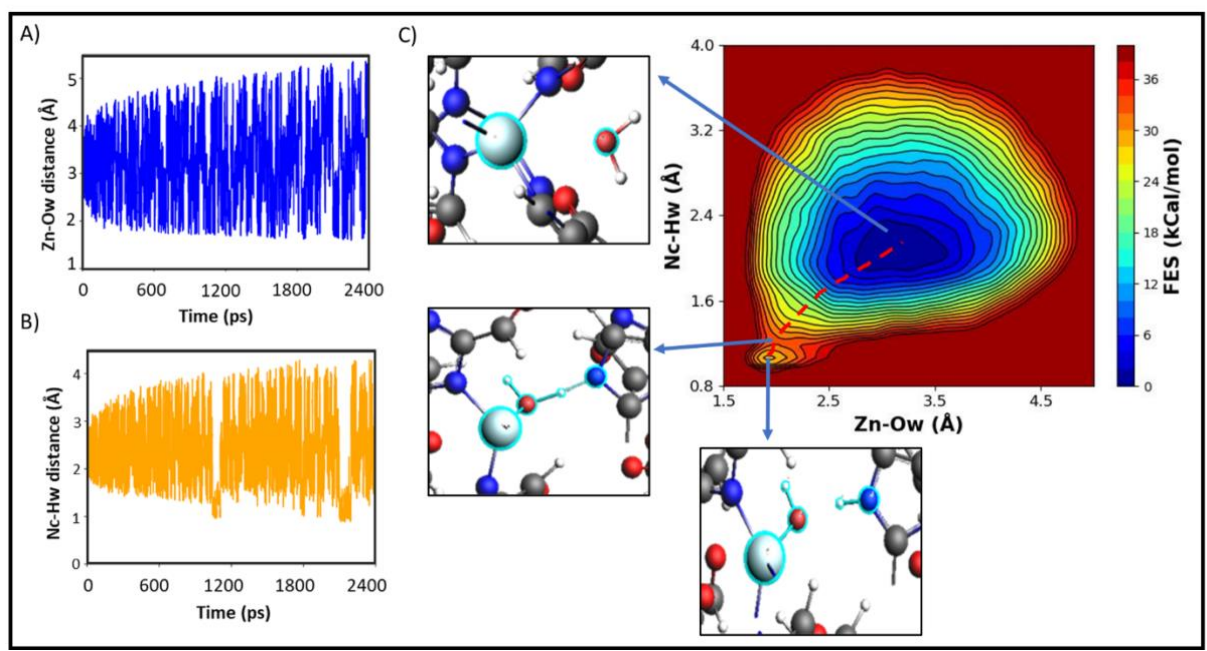

*Figure 3: Progression of the collective variables during a metadynamics simulation of ZIF-90 (Zn-Im(Ica)) with a water molecule at 300 K. The distance from the ZIF metal Zn to the water oxygen (Ow) site is plotted as a function of MD time in (A), The distance from the linker nitrogen (Nc) to the water hydrogen (Hw) is plotted as a function MD time in (B). The FES plot as a function of these two collective variables, constructed from the metadynamics simulation over 2400 picoseconds, illustrating hydrolysis of ZIF-90, is shown in (C). Atomistic representation alongside the FES profile highlighting the reactant to product formation connecting path on the FES. Spheres coloured grey, red, blue and white (small and large) represent C, O, N, H, and Zn respectively.*

The primary energy difference between the two MOFs arises from the breakage of the bond between Zn and the organic linker (Zn-Oc in IRMOF-1 and Zn-Nc in ZIF-90) as well as the protonation of the organic ligand (Oc in IRMOF-1 and Nc in ZIF-90). The higher electronegativity of oxygen compared to nitrogen likely makes the Zn-Oc bond more polarizable, facilitating easier bond breakage in IRMOFs. In contrast, the zinc-nitrogen (Zn-Nc) bond in ZIF is less polarizable and comparatively stronger, making it more challenging to break. This stronger covalent bond in ZIF-90 offers better resistance to reactions with water molecule than in IRMOFs, which, explains the higher barrier for hydrolysis and, consequently, the increased stability. These findings align with the suggested metal-ligand bond strength by Low et al[21] and the strategies reported by Yuan et al. for constructing stable MOFs guided by hard-soft acid-base (HSAB) theory[42]. Therefore, enhancing the robustness of metal-ligand bonds may serve as an effective criterion for identifying water-stable MOFs. Further, we estimated free energy barriers for the hydrolysis of MOF-177 and ZIF-4 with a single water molecule. The barriers are found to be 11±1 kCal/mol and 35±1 kCal/mol, respectively. The detailed progression of the control variables, the reaction mechanism, and the metadynamics estimated free energy surface profile for MOF-177, and ZIF-4 were provided in Figure S3 and Figure S4 of the SI. The estimated hydrolysis free energy barriers align closely with the reported values of 13 kCal/mol for MOF-177 and 40 kCal/mol for ZIF-4 [27].

Next, the influence of chemical descriptors on the free energy barrier is investigated and reported in Figure 4. The functional groups on ligands and their contribution to water stability within MOFs was explored by comparing ZIF-4, which features imidazole as the linker, with ZIF-90 which incorporates 2-imidazole carboxaldehyde as the linker. The presence of the CHO functional group in ZIF-90 likely makes the material more polar and hydrophilic compared to ZIF-4, which lacks this functional group. Consequently, we anticipate ZIF-4 to demonstrate greater stability, characterized by a higher hydrolysis activation-free energy barrier than ZIF-90. Unlike ZIF-90, the frequency of water molecule interacting with zinc in ZIF-4 is lower, resulting in a longer time required to reach reaction states. In line with our expectation, the interaction complex of the water molecule with ZIF-4 is observed at approximately 3.2 Å from the Zn centre, and the hydrolysis of ZIF-4 exhibits a higher activation free energy barrier of 35±1 kCal/mol (Figure S4C). The protonated imidazole exhibits greater stability than the iso-carboxaldehyde, with a free energy barrier of 2 kCal/mol higher than ZIF-90, indicating a higher stability of ZIF-4 than ZIF-90. Hydrolysis is less favourable in ZIF-4, as indicated by a reaction-free energy of 8 kCal/mol (Figure S4C). Also the water adsorption in ZIF-90 is stronger due to the favourable interaction with polar group of iso carboxaldehyde in ZIF-90 than ZIF-4 (Table 2). Moreover, the minimum energy path for the hydrolysis is unique, showing that the dissociation of the water molecule initiates at significantly higher Zn-Ow distances than in previous cases. Moreover, when the imidazolate contains a non-polar functionality, such as methyl groups, the activation-free energy barrier increases substantially to 49.5±3.9 kcal/mol [27]. The detailed analysis highlights that the nature and strength of the metal-ligand bond are pivotal for stability, while additional ligand functional groups provide a modest yet significant influence. This interplay underscores the complexity of stability in metal-organic frameworks.

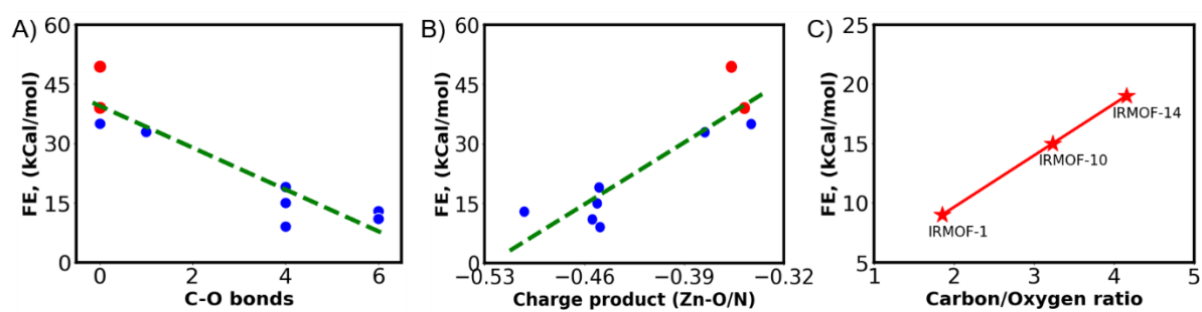

*Figure 4: The free energy barrier is plotted as a function of physicochemical descriptors of the MOFs. The number of C-O bonds, product of Zn and Oxygen or Nitrogen charges, and the ratio of carbon to oxygen atoms are used as a descriptor in (A), (B), and (C), respectively. The coefficient of determination ($R^2$) of the fitted line in (A) and (B) is about ~0.85. The $R^2$ in (C) is 1.0. The two red dots in (A), (B) represent ZIF-7, ZIF-8 from Yang et al.[27].*

As the metal-ligand coordination i.e, number of metal-oxygen bonds increases the oxidation state of metal reduces implying the weakening of Zn-O bond strength in case of MOFs with carboxylic linkers making more vulnerable to react with water. This can be understood by Figure 4A and Figure 4B showing the counter role of C-O bonds in organic ligand and charge product of metal and ligand atom. The carbon-to-oxygen ratio has an increasing trend from IRMOF-1 to IRMOF-10 and IRMOF-14 marking significant repulsive behaviour for water due to higher carbon in organic ligands. This is also evident from the relatively low water adsorption energies of IRMOF-1 and IRMOF-14 (Table-2).

*Table 2: Estimated adsorption energy, activation energy barrier and time scale of MOF-water reaction for all the seven case studies.*

| MOF | Water adsorption energy (kCal/mol) | Energy barrier (kCal/mol) | Estimated reaction time based on energy barrier (hr) |
|---|---|---|---|
| IRMOF-1 | -5.40 | 9 | $1.06 \times 10^{-10}$ |
| IRMOF-10 | -5.40 | 15 | $3.8 \times 10^{-6}$ |
| IRMOF-14 | -4.62 | 19 | $3.1 \times 10^{-3}$ |
| MOF-177 | -6.09 | 11 | $4.6 \times 10^{-9}$ |
| Zn-MOF-74 | -11.70 | 13 | $1.3 \times 10^{-7}$ |
| ZIF-90 | -16.05 | 33 | $4.9 \times 10^{7}$ |
| ZIF-4 | 28.14 | 35 | $1.4 \times 10^{9}$ |

In order to understand the significance of the MOF-water hydrolysis reaction kinetics, it is important to estimate the reaction rates. Here we used Eyring equation $k = \frac{K_B T}{h} e^{-\frac{\Delta G}{RT}}$ [43] to estimate the rate constant of reaction, where k is the reaction rate constant, $K_B$ is the Boltzmann constant, the T is temperature in Kelvin, h is the Planck's constant, ΔG is the Gibbs energy of activation, R is the universal gas constant. Assuming the MOF-water molecule hydrolysis at 300 K as first order reaction step an estimated reaction-time was presented in Table 2 based on the free energy barrier obtained from the metadynamics simulations. It clearly suggests that the ZIF-90 and ZIF-4 are highly stable and will not dissociate in experimental timescale. While IRMOF-1 can degrade in few ns, IRMOR-14 tend to react with a water molecule in a few seconds. The presence of H$_2$O can significantly reduce the CO$_2$ uptake due to the competitive nature of sorbates for the adsorption sites within the pores of the material. This was observed in CALF-20 and ZIF-90[20,44]. In order to use such MOF materials in wet environments for CO$_2$ capture applications a detailed analysis of the fundamental adsorption mechanisms and structural analysis needs to be conducted for all the components in the flue gas mixture.

## 3. Conclusions

We conducted a comprehensive study of water-MOF interactions to uncover mechanistic insights into the hydrolysis reaction of various MOFs using ReaxFF-based metadynamics. By utilizing distances as collective variables, we estimated the free energy landscapes of different MOFs. Based on the energy barrier estimated, all the MOFs studied are stable at room temperature. IRMOFs with smaller pores and lower carbon-to-oxygen ratios are likely less stable than others. In Contrast, ZIFs exhibit a higher energy barrier, indicating greater stability than the IRMOFs. Furthermore, the characteristics of organic ligands, including their carbon-to-oxygen ratios and the functionality of ligands, play a crucial role in the hydrolysis process when water approaches the metal centres. Polar functionalized organic ligand sites can facilitate interactions with water, enhancing reactivity and lowering the free energy barrier. Thus, saturated metal centres linked with non-polar functionalized organic ligands with a higher carbon content than oxygens are anticipated to exhibit greater stability. The combination of weak interactions of water and high energy barrier make the ZIF materials more stable. We infer that ZIFs are promising candidates for CO$_2$/N$_2$ separation in humid conditions, typical of industrial flue gas, and can address the critical challenge of water poisoning that significantly reduces CO$_2$ capture capacity.

**Supporting information**

The file contains additional information on MOFs studied, the ReaxFF Metadynamics estimated free energy surface profile of IRMOF-10, IRMOF-14, MOF-177, ZIF-4, and Zn-MOF-74.

**Acknowledgments**

TKP acknowledges financial support from the SERB, DST, and Govt of India through a core research grant (CRG/2022/006926). JJV acknowledges the Carbon Capture, Utilization and Storage (CCUS) Lab at IIT Madras. AY sincerely thank SLB for providing internship opportunity that contributed to this research.